\documentclass{PoS}

\title{Decaying hadrons within constituent-quark models}

\ShortTitle{Decaying hadrons within constituent-quark models}

\author{R.~Kleinhappel\thanks{Supported by Fonds zur F\"orderung der wissenschaftlichen Forschung in \"Osterreich (FWF DK W1203-N16)}~ and \speaker{W.~Schweiger}\\
        Institut f\"ur Physik, Universit\"at Graz, A-8010 Graz, Austria\\
        E-mail: \email{regina.kleinhappel@uni-graz.at}, \email{wolfgang.schweiger@uni-graz.at}}

\abstract{Within conventional constituent-quark models hadrons come
out as stable bound states of the valence (anti)quarks. Thereby the
resonance character of hadronic excitations is completely ignored. A
more realistic description of hadron spectra can be achieved by
including explicit mesonic degrees of freedom, which couple directly
to the constituent quarks. We will present a coupled-channel
formalism that  describes such hybrid systems in a relativistically
invariant way and allows for the decay of excited hadrons. The
formalism is based on the point-form of relativistic quantum
mechanics. If the confining forces between the (anti)quarks are
described by instantaneous interactions it can be formally shown
that the mass-eigenvalue problem for a system that consists of
dynamical (anti)quarks and mesons reduces to a hadronic eigenvalue
problem in which the eigenstates of the pure confinement problem
(bare hadrons) are coupled via meson loops. The only point where the
quark substructure enters are form factors at the meson-(bare)
hadron vertices. The physical picture that emerges resembles the
kind of hadronic resonance model that has been developed by Sato and
Lee and is now heavily used at the Excited Baryon Analysis Center
(EBAC) to fix $N^\ast$ properties. Our approach, however, is in a
certain sense inverse to the one of Sato and Lee. Whereas they want
to undress physical resonances to end up with bare quantities, we
rather want to dress the bound-states resulting from a pure
constituent quark model to end up with quantities that can be
directly compared with experiment. The way how our approach works
will be exemplified by means of a simple quark-antiquark-meson
system. }

\FullConference{Sixth International Conference on Quarks and Nuclear Physics,\\
        April 16-20, 2012\\
        Ecole Polytechnique, Palaiseau, Paris}

\begin{document}

\section{Introduction}
Although conventional constituent-quark models, in particular their
relativistic versions,  are very successful in reproducing masses
and electroweak properties of the hadron ground states and of the
lowest excited states (see
Refs.~\cite{Capstick:2000qj,Metsch:2008zz,Plessas:2010pk} and
references therein), they are obviously lacking an important piece
of physics. Hadronic excitations result as infinitely long-lived
bound states of the valence (anti)quarks and not as decaying
resonances with a finite lifetime. Strong decays of hadronic
resonances are then usually treated in leading-order perturbation
theory, assuming a particular form for the elementary decay vertex
(e.g. $^3P_0$-model~\cite{Le Yaouanc:1974mr} or elementary emission
model~\cite{Melde:2008yr}) and taking the bound-state wave functions
as resulting from the favorite constituent-quark model for the
incoming and outgoing hadronic states. But it turns out that a large
portion of the calculated partial decay widths underestimate the
experimental values by at least one order of
magnitude~\cite{Metsch:2008zz,Melde:2008yr}. Nevertheless, one can
observe a certain pattern in the decay widths which may be used as
an additional tool for classifying hadronic states according to
$SU(3)$-flavor multiplets~\cite{Melde:2008yr}.

A good starting point for a more realistic description of hadron
resonances  is the physical picture that hadrons consist of a quark
core which is surrounded by a meson cloud. A simple way to realize
this kind of picture within a constituent-quark model is to
introduce, in addition to the valence-(anti)quark
degrees-of-freedom, mesons that can be emitted and absorbed by the
(anti)quarks. Efforts in this direction have been made already (see
Secs. 2.1 and 4.1 of Ref.~\cite{Capstick:2007tv} for an overview),
but mostly within non-relativistic settings. This is certainly
inadequate for hadrons consisting of light quarks as, e.g.,
calculations of the electroweak structure of hadrons clearly
revealed~\cite{Melde:2007zz}. A Poincar\'e invariant formalism for
the description of hybrid systems consisting of valence (anti)quarks
and (elementary) mesons has been proposed in
Ref.~\cite{Krassnigg:2003gh}, where this formalism was applied to
the calculation of vector-meson masses. In the present work we will
follow this approach. It is based on the point-form of relativistic
quantum mechanics and makes use of the Bakamjian-Thomas
construction~\cite{Bakamjian:1953kh} to guarantee Poincar\'e
invariance. But in contrast to Ref.~\cite{Krassnigg:2003gh} we will
also allow for absorption of the emitted meson by the emitting
(anti)quark and not only for the exchange of the meson between
different hadronic constituents.

The relativistic multichannel formalism which we use to treat such
hybrid models that incorporate valence (anti)quarks and mesons is
sketched in Sec.~2. We will show, under the assumption of
instantaneous confining forces, that the corresponding
mass-eigenvalue problem can be equivalently reformulated as a
mass-eigenvalue problem on the hadronic level, in which bare hadrons
are mixed via meson loops. Sec.~3 is devoted to a numerical study of
a very simple system consisting of a quark-antiquark pair and a
meson, all treated as scalar particle. This section contains also
our conclusions and an outlook.

\section{Relativistic coupled-channel framework}
The point-form version of the the Bakamjian-Thomas construction
amounts to the assumption that the (interacting) 4-momentum operator
can be factorized into a free 4-velocity operator and an
interaction-dependent mass operator
\begin{equation}
\hat{P}^\mu = \hat{M} \,\hat{V}^\mu_{\mathrm{free}}
=(\hat{M}_{\mathrm{free}}+\hat{M}_{\mathrm{int}})\,\hat{V}^\mu_{\mathrm{free}}
\, .\label{eq:pfeq}
\end{equation}
Poincar\'e invariance holds if the interaction term
$\hat{M}_{\mathrm{int}}$ is a Lorentz scalar and commutes with
$\hat{V}^\mu_{\mathrm{free}}$. Equation~(\ref{eq:pfeq}) implies that
the overall velocity of the system can be easily separated from the
internal motion which is described by the mass operator $\hat{M}$.
This separation is most conveniently done by representing the
operators in terms of velocity states which specify the state of the
system through its overall velocity and the CM-momenta of its
components~\cite{Klink:1998zz}.

Treating the emission and absorption of particles within such a
quantum-mechanical  framework requires a multichannel formulation.
In the simplest case of only 1 meson $M$ that can be emitted and
absorbed by the valence (anti)quarks one has to deal with a
2-channel problem. The corresponding mass-eigenvalue equation reads
\begin{equation}
 \left(
\begin{array}{cc} {M}_{\mathrm{val}}&  {K}
\\  {K}^\dag&
 {M}_{\mathrm{val},M}
\end{array}\right)
\left(
\begin{array}{l} \vert\Psi_{\mathrm{val}}\rangle
\\ \vert\Psi_{\mathrm{val}, M}\rangle\end{array}
\right) = m \left(
\begin{array}{l} \vert\Psi_{\mathrm{val}}\rangle
\\ \vert\Psi_{\mathrm{val}, M}\rangle\end{array}
\right)\, .\label{eq:meigenv}
\end{equation}
$K^{(\dag)}$ is a vertex operator that accounts for the
absorption (emission) of the meson $M$ by a quark or antiquark. Its
velocity-state matrix elements are obtained from the corresponding
quantum-field theoretical interaction Lagrangian
density~\cite{Klink:2000pp}. The diagonal terms ${M}_{\mathrm{val}}$
and ${M}_{\mathrm{val},M}$ of the matrix mass operator $\hat{M}$ contain,
in addition to the relativistic kinetic energies of the valence
(anti)quarks and the meson $M$, an instantaneous confining potential
between the (anti)quarks and, possibly, also an instantaneous
hyperfine interaction. As it turns out in the following, the
assumption that confinement is instantaneous will be crucial for the
physical interpretation of this kind of model. Physical hadrons,
i.e. eigenstates of $\hat{M}$, are then a superposition of
$\vert\Psi_{\mathrm{val}}\rangle$ and $\vert\Psi_{\mathrm{val},
M}\rangle$. In order to solve the mass-eigenvalue
equation~(\ref{eq:meigenv}) we, however, eliminate
$\vert\Psi_{\mathrm{val}, M}\rangle$ to end up with an equation for
$\vert\Psi_{\mathrm{val}}\rangle$ alone:
\begin{equation}\label{eq:feshbach}
\big(M_{\mathrm{val}}+ {\underbrace{{ {K}\big(m -
{M_{\mathrm{val},M}} \,\,\,\, \big)^{-1} {K}^{\dag}}}_{={
{V}_{\mathrm{opt}}}\left( m\right)
}}\big)\vert\Psi_{\mathrm{val}}\rangle =m\,
\vert\Psi_{\mathrm{val}}\rangle\, .
\end{equation}
${V}_{\mathrm{opt}}$ is an optical potential that describes the
emission of the meson $M$ by a valence (anti)quark and the
subsequent absorption by another or the same valence (anti)quark.
The important point to note here is that the system which propagates in the intermediate state consists of a meson and confined valence (anti)quarks. Therefore the valence (anti)quarks are not renormalized by meson loops.
Equation~(\ref{eq:feshbach}) can now be further reduced to an
algebraic equation by expanding $\vert\Psi_{\mathrm{val}}\rangle$ in
terms of eigenstates of $M_{\mathrm{val}}$:
\begin{equation}
\vert \psi_{\mathrm{val}}\rangle=\sum_n\, A_n\, \vert n\rangle
\quad\hbox{with}\quad {M}_{\mathrm{val}}\vert n\rangle =
\mu_n \vert n \rangle\, .
\end{equation}
We will call the eigenstates $\vert n \rangle$ of $M_{\mathrm{val}}$ \lq\lq bare hadrons\rq\rq. A physical hadron is thus a superposition of bare hadrons with the coefficients $A_n$ being given by the set of algebraic equations
\begin{equation}\label{eq:hadronev}
(m-\mu_n)\, A_n =  \sum_{n^{\prime}}\, \langle n \vert
\,{V}_{\mathrm{opt}}(m{ + i\epsilon})\, \vert n^{\prime} \rangle\,
\, A_{n^{\prime}}\, , \qquad n=0,1,2,3,\dots\, .
\end{equation}
\begin{figure}[h!]
\includegraphics[width=.48\textwidth]{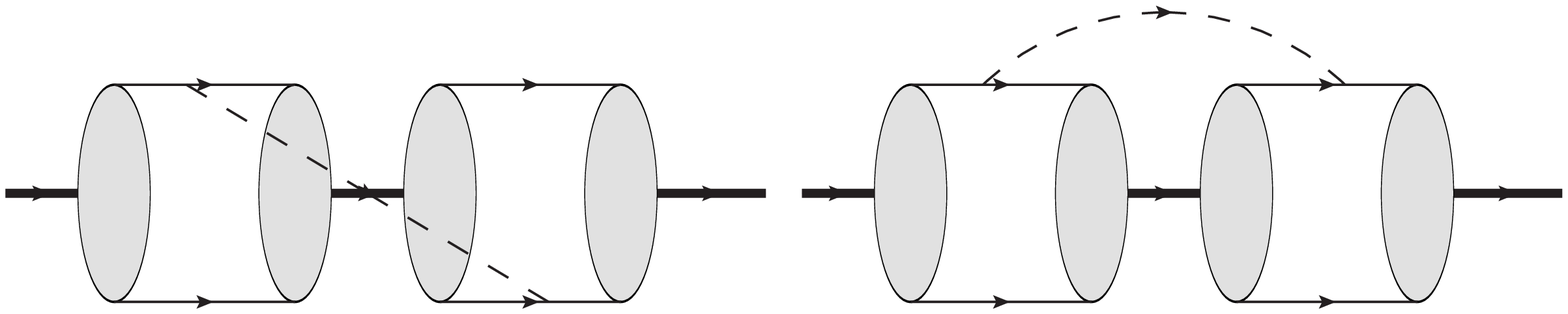}
\includegraphics[width=.48\textwidth]{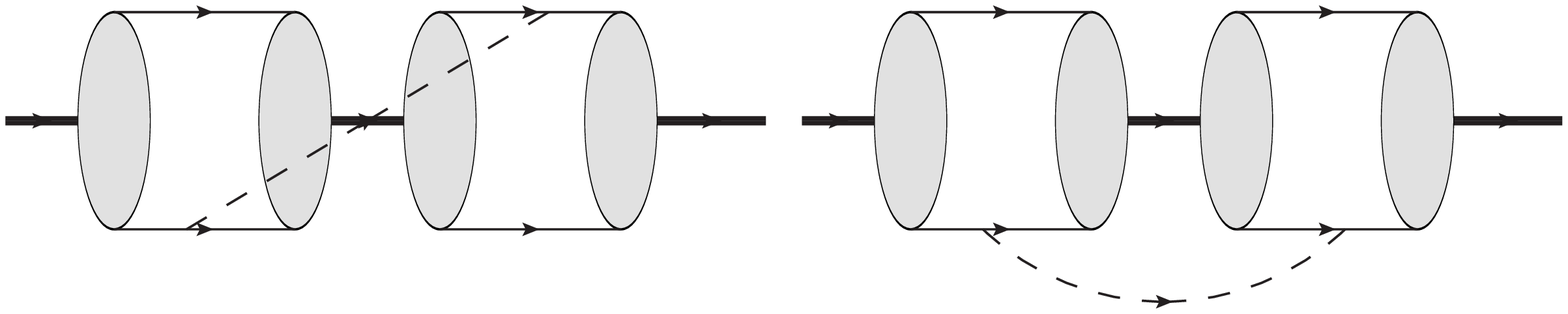}\\
\vspace{-0.8cm}
$$
\Updownarrow
$$
\\
\vspace{-1.3cm}
\begin{center}
\includegraphics[width=4cm]{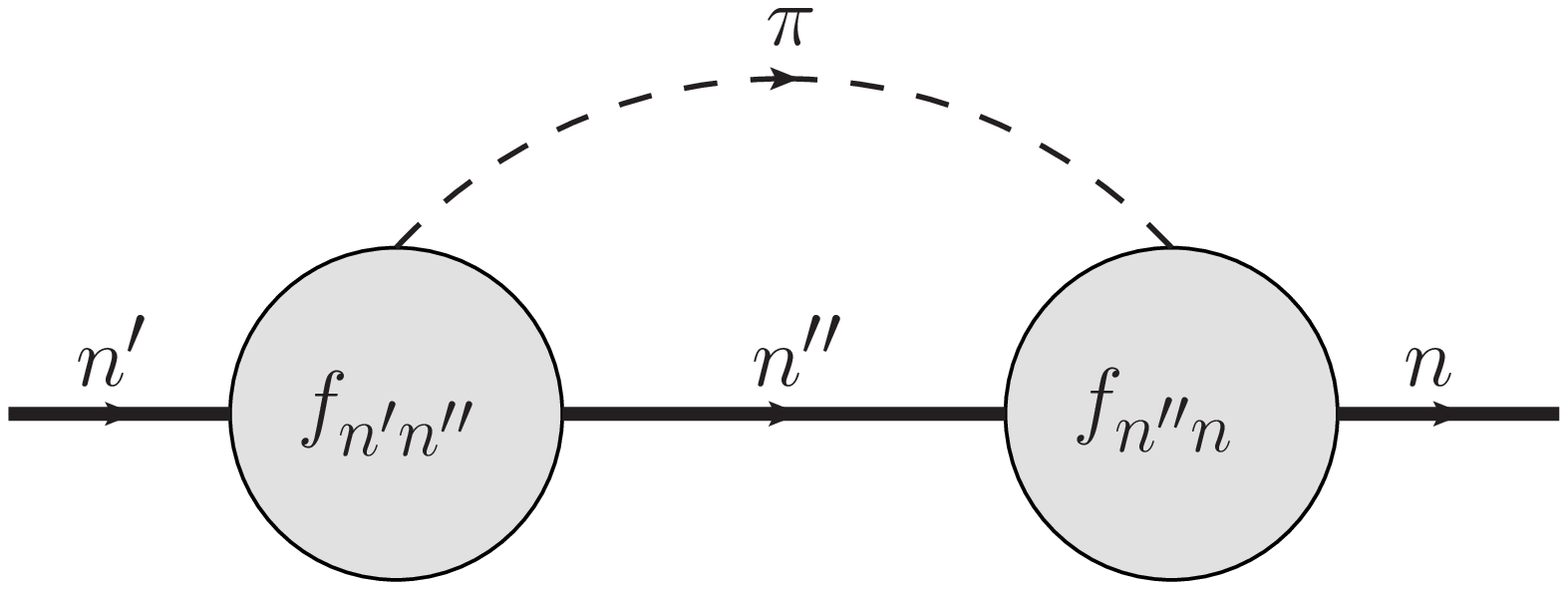}
\end{center}
\caption{Graphical representation of the optical potential showing up in Eq.~(2.5) for a confined quark-antiquark system that is allowed to emit and reabsorb a pion. The optical potential on the constituent level is represented by the upper graphs, the equivalent optical potential on the hadronic level by the lower graph.}\label{fig:vopt}
\end{figure}
This system of equations is now already an eigenvalue problem on the hadronic level that describes bare hadrons which are coupled via meson loops. A further analysis of the matrix elements of the optical potential reveals indeed that all the valence-(anti)quark substructure of the bare hadrons can be absorbed into strong meson-(bare) hadron vertex form factors $f_{n,n^{\prime}}(|\vec{k}_M|)$ (see Fig.~\ref{fig:vopt}) such that our model acquires an equivalent formulation on the hadronic level. It is now important to note that the optical potential becomes singular as soon as $m>\mu_0+m_M$, i.e. the mass eigenvalue $m$ becomes larger than the mass $\mu_0$ of the bare ground state $|n=0\rangle$ plus the meson mass $m_M$. This singularities are treated by means of the usual $i\epsilon$-prescription. As a consequence, the matrix elements of the optical potential $\langle n \vert {V}_{\mathrm{opt}}(m{ + i\epsilon}) \vert n^{\prime} \rangle$ become complex. Most importantly, this leads to complex mass eigenvalues ${\textbf m}_j=m_j+ i\Gamma_j/2$ and thus to finite decay widths $\Gamma_j$ for excited states ($j>0$). Practically we have solved the eigenvalue equation~(\ref{eq:hadronev}) in which the mass eigenvalue $m$ enters in a non-linear way (via $V_\mathrm{opt}(m)$) by means of an iterative solution method which provides also the complex mass eigenvalues~\cite{Kleinhappel:2010}.

\section{Numerical results and outlook}
First numerical studies along these lines were performed in Ref.~\cite{Kleinhappel:2010} for a simple model of meson resonances. The degrees-of-freedom of this model are $1$ quark, $1$ antiquark and 1 pion that can be emitted and absorbed by the (anti)quark. Spin and flavor are neglected in this model and confinement of the quark-antiquark pair is described via a harmonic-oscillator potential (added, however, to $M^2_{q\bar{q},\,\mathrm{free}}$). The mass parameters of this model were prefixed to commonly used values ($m_q=m_{\bar{q}}=0.34$~GeV, $m_\pi=0.135$~GeV). The $\pi-q$ coupling constant $g$ was varied within a reasonable range to study its effect on the mass eigenvalues and the decay widths. To give the model some physical meaning the parameters of the confinement potential (oscillator strength $a=0.28$~GeV and parameter $V_0=0.1$~GeV that fixes the ground-state mass) were determined a posteriori such that the $\omega(782)$ and $\omega(1420)$ masses are approximately reproduced for the maximum decay width $\Gamma_1$ that can be achieved. The study was restricted to $s$-wave mesons and orbital excitations were neglected in intermediate states.

\begin{figure}
\includegraphics[width=.48\textwidth]{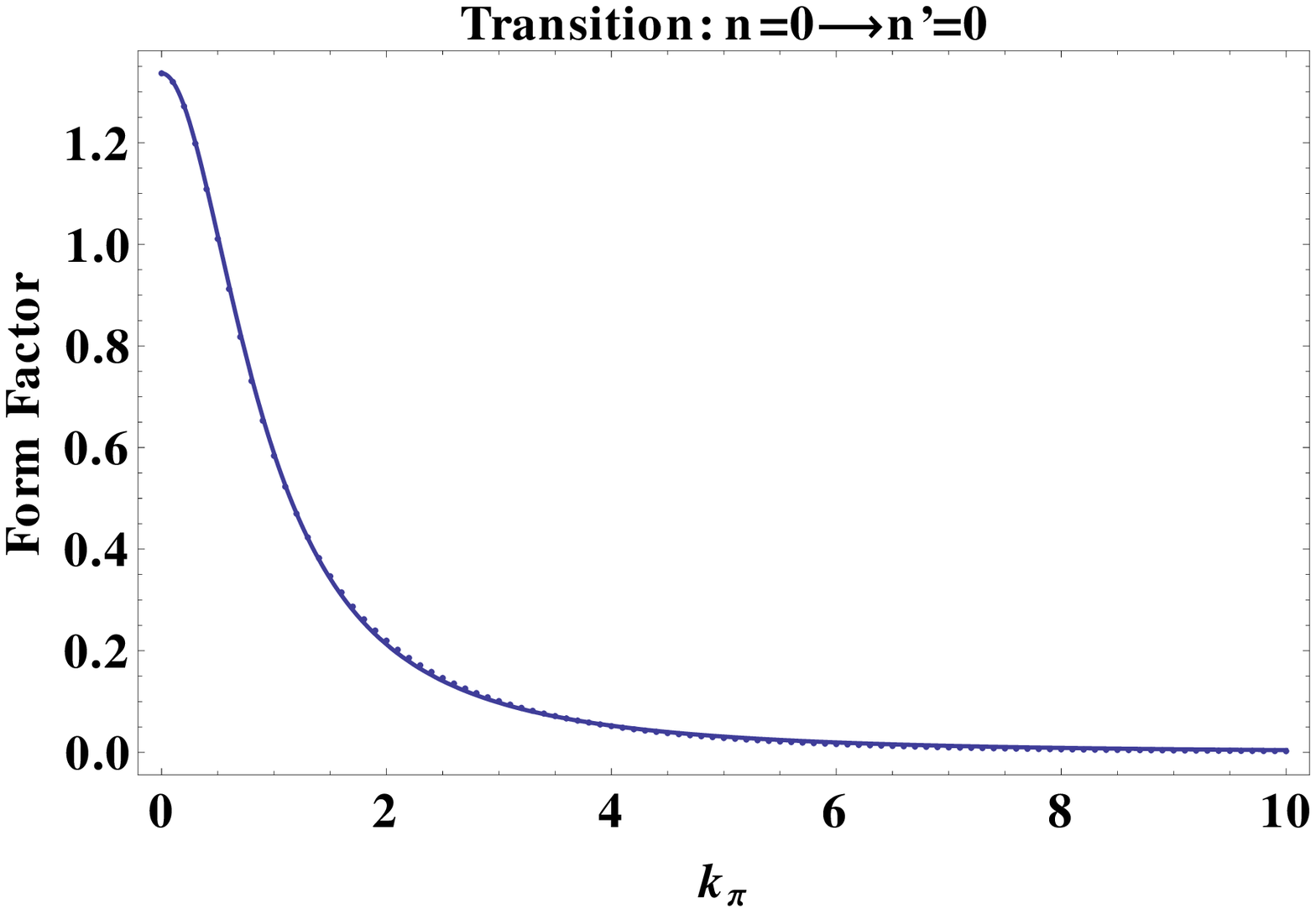}\hspace{0.3cm}
\includegraphics[width=.48\textwidth]{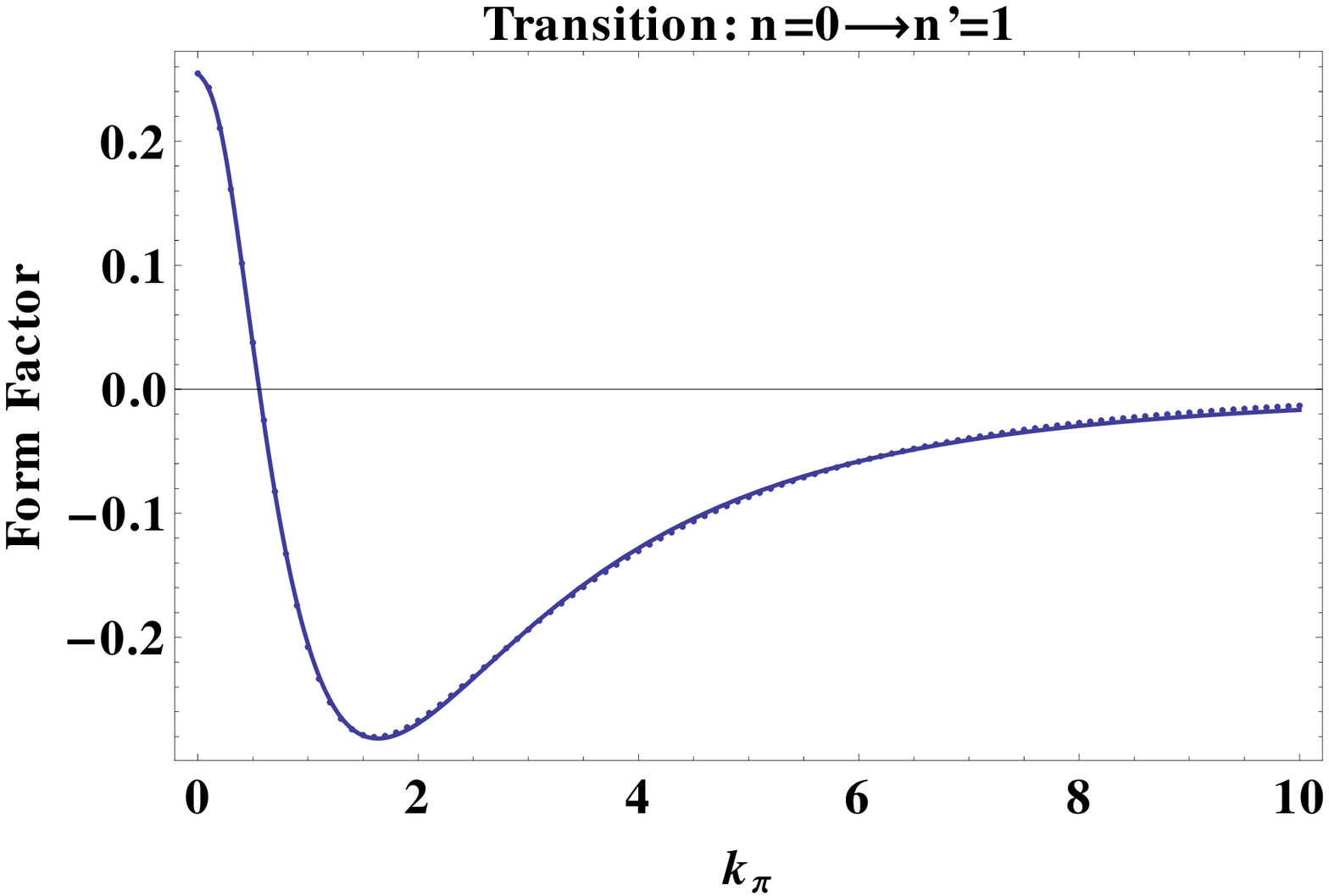}
\caption{Left: elastic form factor for the coupling of a pion to the
harmonic-oscillator ground state ($n=n^\prime=0$). Right: transition
form factor associated with  pion emission causing a transition of
the $n = 0$ to the $n^\prime = 1$ harmonic oscillator state. Form
factors are plotted as functions of $|\vec{k}_\pi|$ and are not
normalized.} \label{fig:ffs}
\end{figure}

\begin{figure}\begin{center}
\includegraphics[width=.6\textwidth]{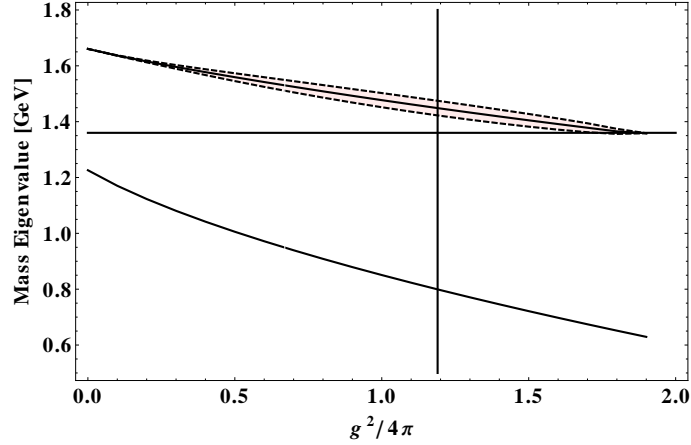}\hspace{1cm}\end{center}
\caption{Dependence of the ground state and the first excited state
on the $\pi$-$q$ coupling constant $g$. The shaded band indicates the decay
width $\Gamma_1$ of the first excited state (multiplied by $4$ for
better visibility). The positions of the maximal decay width ($\Gamma_1=0.026$~GeV reached at
$g^2/4\pi=1.19$~GeV$^2$) and of the decay threshold of the first excited
state (at $\mu_0+m_\pi$) are indicated by vertical and horizontal
lines, respectively.} \label{fig:spectrum}
\end{figure}

The first physical information one can extract from this model is the structure of the pion-(bare)-meson vertices which is encoded in the (strong) form factors $f_{n,n^{\prime}}(|\vec{k}_\pi|)$. Two of these form factors are plotted in Fig.~\ref{fig:ffs}. The dependence of the ground-state mass and the mass and width of the first excited state on the $\pi-q$ coupling constant $g$ is shown in Fig.~\ref{fig:spectrum}. The $\pi$-loop obviously provides an attractive force which pushes the masses downwards. The decay width of the first excited state $\Gamma_1$ exhibits a maximum as function of $g$ and vanishes when ${\mathrm{Re}}(\textbf{m}_{1})=\mu_0+m_\pi$. This behavior indicates that our simple model, although it provides hadron excitations with finite lifetimes, is still not complete. Instead of decaying into the physical ground state (mass $\textbf{m}_{0}$) and a $\pi$ the first excited state can only decay into the bare ground state (mass $\mu_0$) and a $\pi$. A way out would be to increase the number of channels and include $q\bar{q}2\pi$, $q\bar{q}3\pi$, etc. channels. This has do be done until the probability of finding $n$ pions in the (physical) meson becomes negligibly small. The technical problems associated with such a procedure would soon become rather big, maybe even unsurmountable.

Another, more promising strategy would be to consider the present calculation just as a first step to \lq\lq dress\rq\rq\ bare hadrons (i.e. the confined valence-(anti)quark states) by means of meson loops. The important point to notice is that in our approach this dressing happens already on the hadronic level and not on the quark level if confinement is represented by an instantaneous interaction. In a next step the meson-(bare)hadron vertices and the masses of the bare hadrons, coming from the hybrid quark model, could be used as the basic input for a dynamical coupled-channel calculation of hadron masses and decay widths à la Sato and Lee~\cite{Sato:2009}. In this second step all the calculations are already done on the hadronic level. Put into action, such a bottom-up approach could provide an understanding of hadron resonances in terms of constituent-quark degrees-of-freedom that goes beyond pure hadron reaction models à la Sato and Lee and it would substantiate the physical picture of hadrons consisting of a quark core that is surrounded by a meson cloud.

\end{document}